\documentstyle[twocolumn,seceq]{jpsj}
\input epsf  

\def\runtitle{Symmetry of high-piezoelectric Pb based complex 
perovskites at the morphotropic . . . .}
\def\runauthor{Yoshiaki {\sc Uesu} {\it et al}.}

\title
{
Symmetry of high-piezoelectric Pb based complex perovskites 
at the morphotropic phase boundary\\ I. Neutron diffraction 
study on Pb(Zn$_{1/3}$Nb$_{2/3}$)O$_{3}$-9$\%$PbTiO$_{3}$ 
}

\author
{ 
Yoshiaki {\sc Uesu}$^{1,2,}$\footnote{E-mail: uesu93@mn.waseda.ac.jp}, 
Masaaki {\sc Matsuda}$^{3}$, 
Yasusada {\sc Yamada}$^{2,3}$, 
Kouji {\sc Fujishiro}$^{4}$, 
Dave E. {\sc Cox}$^{5}$, 
Beatriz {\sc Noheda}$^{5}$ 
and Gen {\sc Shirane}$^{5}$
}

\inst
{
$^1$Department of Physics, Waseda University, 3-4-1 Okubo, Shinjuku-ku, 
Tokyo 169-8555, Japan\\
$^2$Advanced Research Institute, Waseda University, 3-4-1 Okubo, Shinjuku-ku, 
Tokyo 169-8555, Japan\\
$^3$Advanced Science Research Center, Japan Atomic Energy Research Institute, 
Tokai, Ibaraki 319-1195, Japan\\
$^4$Materials and Structures Laboratory, Tokyo Institute of Technology, 
4259 Nagatsuda, Midori-ku, Yokohama 226-8503, Japan\\
$^5$Department of Physics, Brookhaven National Laboratory, Upton, 
New York 11973, USA
}

\recdate
{
\today
}

\abst
{
The symmetry was examined using neutron diffraction method on 
Pb(Zn$_{1/3}$Nb$_{2/3}$)O$_{3}$-9$\%$PbTiO$_{3}$ (PZN/9PT) which 
has a composition at the morphotropic phase boundary between 
Pb(Zn$_{1/3}$Nb$_{2/3}$)O$_{3}$ and PbTiO$_{3}$. The results were 
compared with those of other specimens with same composition but with 
different prehistory. The equilibrium state of all examined specimens 
is not the mixture of rhombohedral and tetragonal phases of the end members 
but exists in a new polarization rotation line M$_{\mathrm{c}}^{\#}$ 
(orthorhombic-monoclinic line). Among examined specimens, one exhibited 
tetragonal symmetry at room temperature but recovered monoclinic phase 
after a cooling and heating cycle. 
}

\kword
{
Piezoelectricity, ferroelectricity, morphotropic phase boundary, 
neutron diffraction, Pb(Zn$_{1/3}$Nb$_{2/3}$)O$_{3}$-9$\%$PbTiO$_{3}$
}

\begin{document}
\sloppy
\maketitle

\section{Introduction}
Extremely high piezoelectricity was found near the morphotropic phase 
boundary (MPB) of mixture compounds of Pb-based complex perovskite oxides 
and PbTiO$_{3}$.~\cite{rf:1} 
Examples are 
92$\%$Pb(Zn$_{1/3}$Nb$_{2/3}$)O$_{3}$-8$\%$PbTiO$_{3}$~\cite{rf:2}, \ \
52$\%$PbZrO$_{3}$-48$\%$PbTiO$_{3}$~\cite{rf:3} \ \
and 65$\%$Pb(Mg$_{1/3}$Nb$_{2/3}$)O$_{3}$-35$\%$PbTiO$_{3}$~\cite{rf:1}.
Hereafter they are abbreviated as PZN-$x$PT, PZT$x$ and PMN-$x$PT with 
composition ratio $x(\%)$ of PbTiO$_{3}$.
In fact the electro-mechanical coupling coefficient $k_{33}$ exceeds 90\% 
at the MPB where a strain up to 1.7\% can be induced by an electric field.
Furthermore some materials exhibit low hysteresis strain-field curve. 
These properties have attracted much attention to the piezoelectric 
applications, i.e., energy-conserving transformation of electric and 
mechanical energy into one another in medical diagnostic apparatus, 
actuators, high power ultrasonic transducers, underwater acoustic and so on.

The origin of ultra-high piezoelectricity at the MPB of these materials has 
been puzzling for long time. However recent two approaches promoted a better 
understanding of the phenomenon: One is the first-principle study by Fu and 
Cohen.~\cite{rf:4}
Under the assumption that a large piezoelectricity is driven by polarization 
rotation induced by an electric field, they calculated the most favorable path 
in BaTiO$_{3}$, which is similar to the high-piezoelectric complex systems. 
Observed strain-field relation was quantitatively explained by choosing a path 
where polarization rotates between two extremities of rhomboheral [111] and 
tetragonal [001] axes in the monoclinic plane (M$_{\mathrm{a}}$).
Later the existence of monoclinic phases was proved to be compatible with 
the Devonshire theory by introducing higher order term P$^8$ which reflects 
strong unharmonicity of the system.~\cite{rf:5}
The first principle calculation also showed that the high-piezoelectricity 
is induced due to the existence of a monoclinic phase.~\cite{rf:6}

\begin{figure}[htbp]
\centering{\epsfxsize 3.5cm \epsfbox{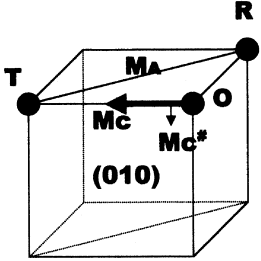}}
\caption{
Possible polarization rotation paths in high-piezoelectric complex 
compounds at the morphotropic phase boundary. T, O and R indicate 
the polarization directions in tetragonal, orthorhombic and 
rhombohedral phases, respectively. M$_{\mathrm{a}}$ and 
M$_{\mathrm{c}}$ are polarization rotation paths in monoclinic 
(1$\bar{1}$0) and (010) planes, respectively. M$_{\mathrm{c}}^{\#}$ is 
a new polarization rotation path found in PZN/9PT.
}
\label{fig:1}
\end{figure}

Another approach to the origin of high-piezoelectricity is experimental. 
Noheda et al. examined precisely crystal symmetries at a very narrow region 
of the MPB of PZT48, and found that the PZT48 shows monoclinic structure 
with the point group Cm with polarization vector lying in the monoclinic 
plane M$_{\mathrm{a}}$.~\cite{rf:7,rf:8,rf:9,rf:10}
They also performed measurements on PZN-8PT samples and found that PZN-8PT 
has rhombohedral symmetry in as-grown state but an orthorhombic phase is 
irreversibly induced by the electric field.~\cite{rf:11}
It indicates that examined samples are in the proximity of a 
rhombohedral-orthorhombic boundary in the phase diagram. 
In fact, recent high-resolution synchrotron X-ray diffraction study by 
Cox et al. revealed that a PZN-9PT powder sample (9PT-2) prepared carefully 
from a poled single crystal shows an orthorhombic symmetry.~\cite{rf:12}
Orthorhombic phase is another extremity in polarization rotation path 
(M$_{\mathrm{c}}$) connecting [001] to [101] as shown in Fig.~\ref{fig:1}.
In this case, the polarization possibly rotates within the (010) 
monoclinic plane of the point group Pm under the external electric 
along the [001] direction. As a necessary condition for high-piezoelectricity 
is the existence of monoclinic phase,~\cite{rf:4, rf:5} PZN-9PT would 
exhibit also a monoclinic phase depending on the history of sample treatment.
For this motivation, we examined four PZN-9PT samples including 9PT-2 
in order to determine the symmetry at the MPB of PZN-PT system. 
In particular, measurements of two samples (9PT-4 and 9PT-5) were 
performed using neutron diffraction method.

\section{Experimental conditions}
PZN-9PT single crystals were grown by a flux method.~\cite{rf:13} 
Four samples were prepared with different sample prehistory. 
Among them 9PT-1 is single crystal plates cut from same as-grown 
single crystal, 9PT-2 a powder sample prepared from a poled single 
crystal, 9PT-4 a (001) plate which has been once exposed to the 
electric field, and 9PT-5 a (001) plate which has never experienced 
the field, similar to 9PT-1, but cut from a different single crystal. 

Neutron diffraction measurements were performed on 9PT-4 and 5 using 
triple-axis spectrometers TAS1 and TAS2 installed in JRR-3M in Japan 
Atomic Energy Research Institute. For the measurements, the horizontal 
collimator sequences were 20'-20'-S-20'-20' with Ei=13.64meV 
($\lambda$=2.449\AA) 
on TAS1 and guide-20'-S-20'-10' with Ei=13.72meV ($\lambda$=2.441\AA) 
on TAS2. Diffraction profiles of 9PT-5 were measured by a mesh-scanning 
method in the \{001\}$^*$ 
plane at room temperature, 10K, 80K and 200K. Measurements of 9PT-4 and of 
after-cooling profiles of 9PT-5 were performed by a simple scanning method.

\section{Results and discussions}
\subsection{Neutron diffraction study on 9PT-5}
\begin{figure}
\centering{\epsfxsize 7.5cm \epsfbox{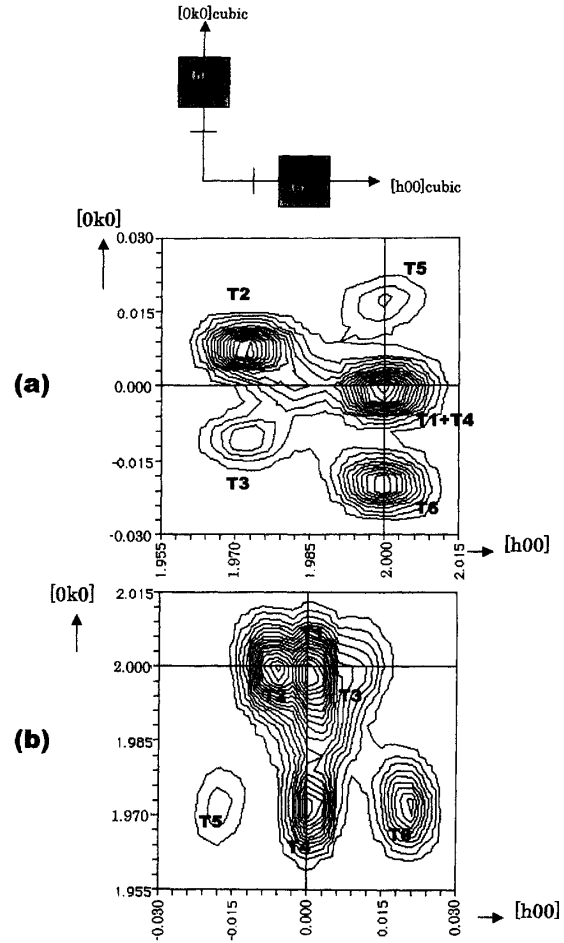}}
\caption{
Neutron diffraction profiles at room temperature around 020 
reciprocal point in the (hk0) (a) and 200 in the (h0l) planes 
of 9PT-5. They are a juxtaposition of tetragonal domain 
configurations T$_i$ ($i$=1 to 7) as shown in Fig.~\ref{fig:3}(a). 
The referred axes are those of cubic phase, and the lattice point 
with h(or k)=2 corresponds to Q=3.116\AA$^{-1}$. 
The upper figure shows scanning areas of the present experiment. 
}
\label{fig:2}
\end{figure}

\begin{figure}
\centering{\epsfxsize 7cm \epsfbox{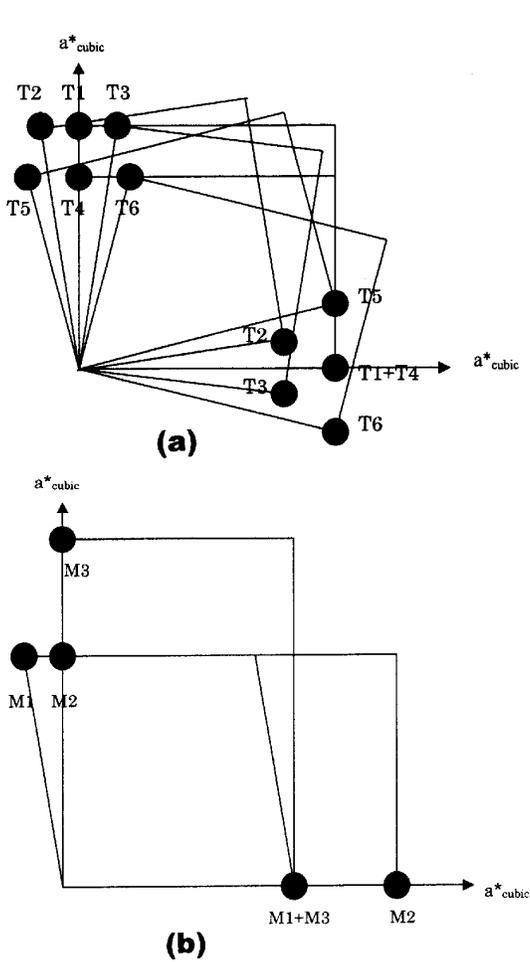}}
\caption{
(a) indicates tetragonal domain configuration corresponding 
to the observed pattern in Fig.~\ref{fig:2}, 
and (b) monoclinic one corresponding to Fig.~\ref{fig:4}. 
Solid circles represent lattice points of these domains
}
\label{fig:3}
\end{figure}

Diffraction profiles in (hk0) and (h0l) planes were measured at room 
temperature.
Fig.~\ref{fig:2} shows those around 200 (a) and 020 (b) reciprocal 
points observed in the (hk0) plane. 
Here the ordinate and abscissa are referred to the cubic axes and 
the lattice point with h(k)=2 corresponds to Q=3.116\AA$^{-1}$. 
Several peaks are observed and can be interpreted as a juxtaposition 
of tetragonal domain configurations, which are indicated 
in Fig.~\ref{fig:3}(a). 
Seven domains T$_i$ ($i$=1 to 7) existed in an irradiated part of the sample.
In particular, from
the locations of T$_1$ and T$_4$, tetragonal lattice 
parameters were determined as a=c=4.033\AA, b=4.085\AA. 
These values coincide well with those of 9PT-1, which is a mixture of 
orthorhombic and tetragonal phases, as will be discussed later. 

\begin{figure}
\centering{\epsfxsize 7.5cm \epsfbox{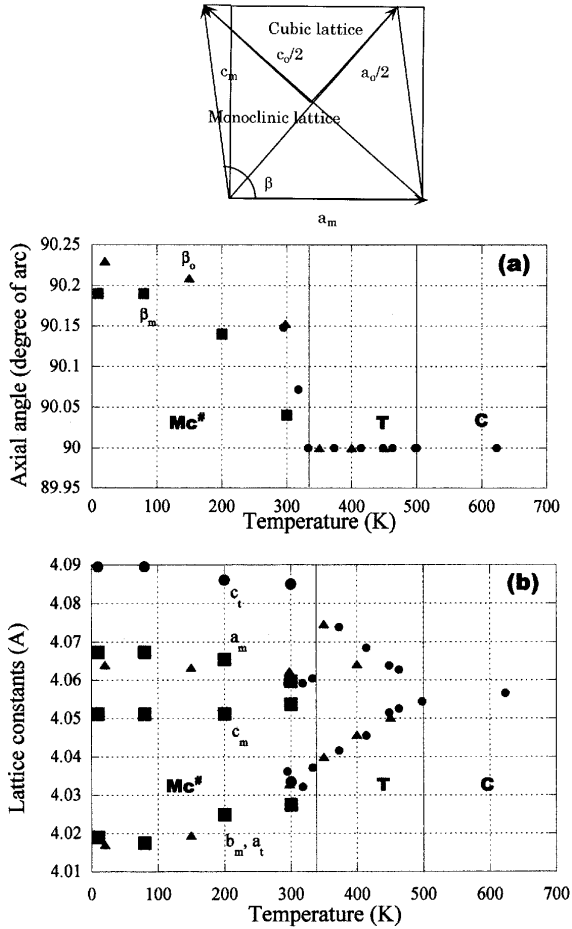}}
\caption{
Temperature dependences of monoclinic lattice parameters 
a$_{\mathrm{m}}$, b$_{\mathrm{m}}$, c$_{\mathrm{m}}$ and 
$\beta$$_{\mathrm{m}}$ of 9PT-5. 
Orthorhombic lattice parameters a$_{\mathrm{o}}$=c$_{\mathrm{o}}$, 
b$_{\mathrm{o}}$ and $\beta$$_{\mathrm{o}}$ 
of 9PT-1 and 2~\cite{rf:15} are shown for comparison. 
The axes are referred to the cubic ones. 
(a) indicates axial angles $\beta$, and (b) lattice constants
a,b and c. 
Here red squeres indicate monoclinic lattice parameters 
a$_{\mathrm{m}}$, b$_{\mathrm{m}}$, c$_{\mathrm{m}}$, 
$\beta_{\mathrm{m}}$ of 9PT-5, 
blue circles tetragonal lattice parameters a$_{\mathrm{t}}$=b$_{\mathrm{t}}$, 
c$_{\mathrm{t}}$ 
of 9PT-5, violet triangles of 9PT-2, and solid circles of 9PT-1.
}
\label{fig:5}
\end{figure}

The tetragonal phase could appear even at room temperature as a quenched 
high-temperature phase. Therefore if the sample is cooled down to a low 
temperature, it would be transformed into the ground state, which would 
be maintained after heating. In this expectation, we performed low 
temperature measurements with same sample and similar diffraction geometry.
Fig.~\ref{fig:3} shows diffraction profiles observed at 10K. 
Measurements were made around same reciprocal points as in Fig.~\ref{fig:2}.
\begin{figure}
\centering{\epsfxsize 7.5cm \epsfbox{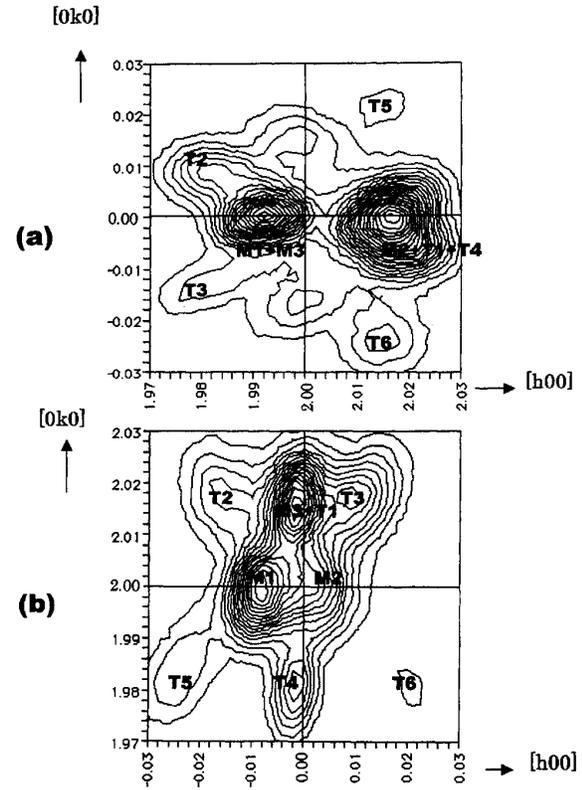}}
\caption{
Neutron diffraction profiles at10K around 020 (a) and 200 (b) 
reciprocal points in the (hk0) plane of 9PT-5. M$_{1}$, M$_{2}$ and M$_{3}$ 
indicate corresponding monoclinic domains in Fig.~\ref{fig:3}(b). 
The referred axes are those of cubic and the lattice point h=2 on 
the [h00] axis corresponds to Q=3.102\AA$^{-1}$, and k=2 on the [0k0] 
axis to Q=3.090\AA$^{-1}$.
}
\label{fig:4}
\end{figure}
Here (a) indicates those around 020 and (b) 200 both in the (hk0) plane. 
Contrary to the profiles at room temperature, main peaks are interpreted 
by monoclinic domains. A small portion of tetragonal domains still exist 
in low temperature but with a marked diminution of intensities, e.g., 
that of T$_6$. Three types of monoclinic domain M$_1$, M$_2$ and M$_3$ 
contribute to the profiles as shown in Fig.~\ref{fig:4}. 
Lattice parameters were determined as a=4.051\AA, b=4.019\AA, c=4.067\AA, 
$\beta$=90.19$^{\circ}$ in monoclinic phase, and a=b=4.019\AA, c=4.090$\AA$ 
in tetragonal phase.
Similar measurements were performed at 80K, 200K and room temperature in 
this sequence. Essentially similar profiles were observed at these 
temperatures and the monoclinic symmetry is maintained in higher 
temperature. 
Lattice parameters determined at 10K and RT are tabulated 
in Table~\ref{table:1}, 
and their temperature dependences 
are shown in Fig.~\ref{fig:5}(a) for axial angle $\beta$ and (b) 
for lattice constants, where those of 9PT-1 and 2 are shown 
for comparison. 
Here $\beta$ means a monoclinic axial angle between 
a$_{\mathrm{m}}$ and c$_{\mathrm{m}}$ as shown in the upper 
figure of Fig.~\ref{fig:5}(a). When a$_{\mathrm{m}}$=c$_{\mathrm{m}}$, 
the crystal system becomes orthorhombic with a$_{\mathrm{o}}$ and 
c$_{\mathrm{o}}$ which are perpendicular to each 
other (Fig.~\ref{fig:5}(a)). 
However, the monoclinic axes are chosen in 
Table~\ref{table:1} and Fig.~\ref{fig:5}, $\beta$ of the orthorhombic 
limit being indicated by $\beta_{\mathrm{o}}$. 
The monoclinic lattice parameters vary with temperature almost in same 
manner as orthorhombic ones. This means that the observed monoclinic 
state is located near the orthorhombic limit in the (010) plane 
(Fig.~\ref{fig:1}). 
\begin{fulltable}
\caption{Lattice parameters of 9PT-1, 2, 4 and 5.}
\label{table:1}
\begin{fulltabular}{@{\hspace{\tabcolsep}\extracolsep{\fill}}ccccccc} \hline
 & 9PT-1 & 9PT-2 & 9PT-4 & \ \ & 9PT-5 & \ \ \\ \hline
Single crystal or powder & Single crystal & Powder$^{\mathrm{a,}}$~\cite{rf:12} & Single crystal & \ \ & Single crystal & \ \ \\
Prehistory of field application	& No & Yes & Yes & \ \ & No & \ \ \\
Temperature & RT & RT & RT & RT$^{\mathrm{b}}$ & 10K & RT$^{\mathrm{c}}$ \\
Symmetry & O $+$ T & O & O & T & M$^{\mathrm{d}}$ $+$ T & M $+$ T \\
Lattice parameters & O: & & & & M: & M: \\
&a=4.059&a=4.0623&a=4.062&a=4.033&a=4.051&a=4.052\\
&b=4.036&b=4.0328&b=4.033&b=4.033&b=4.019&b=4.028\\
&c=4.059&c=4.0623&c=4.062&c=4.085&c=4.067&c=4.057\\
&$\beta_{\mathrm{o}}$=90.15&$\beta_{\mathrm{o}}$=90.15&$\beta_{\mathrm{o}}$=90.15&&$\beta$=90.19&$\beta$=90.11\\
&T:&&&&T:&T:\\
&a=4.0317&&&&a=4.019&a=4.028\\
&b=4.0317&&&&b=4.019&b=4.028\\
&c=4.0840&&&&c=4.090&c=4.078\\
Experimental method& X-ray & SR & Neutron & & Neutron & \ \ \\ \hline
\end{fulltabular}
$^{\mathrm{a}}$Prepared from a poled single crystal.
$^{\mathrm{b}}$Before cooling.
$^{\mathrm{c}}$After cooling.
$^{\mathrm{d}}$M is dominant.
\end{fulltable}

\subsection{Re-examination of the symmetry of 9PT-1}
On the procedure of systematic analyses of the symmetry of PZN/9PT, 
we noticed that the previously published data of 9PT-1~\cite{rf:14} 
should be re-examined. 
$\theta$-2$\theta$ scanning profiles were measured using 
(111) and (100) plates 
in temperature range from 300 to 625K by a high-accuracy X-ray 
diffractometer in SPMS, Ecole Centrale Paris. (222) and (400) profiles 
were shown in Fig.~\ref{fig:6}.
They have been analyzed under an assumption that rhombohedral and tetragonal 
phases co-exist below the ferroelectric phase transition at 500K. 
In this analysis, tetragonal lattice constants a, b, c and rhombohedral 
a(=b=c) were determined independently from the (100) plate and (111) 
plate experiments, respectively. However, we encountered a discrepancy 
when we tried to explain the (400) profile by the mixture of tetragonal 
and rhombohedral reflections which were determined from the (222) 
reflection as shown in Fig.~\ref{fig:6}(a). 
In particular, the peak around 2$\theta$=86.6$+{\circ}$ cannot be 
explained 
by the model. 
The problem could be solved when orthorhombic symmetry is considered 
to be true in place of rhombohedral one (Fig.~\ref{fig:6}(b)). 
The determined lattice constants coincide well with those of 9PT-2.
\begin{fullfigure}
\centering{\epsfxsize 13cm \epsfbox{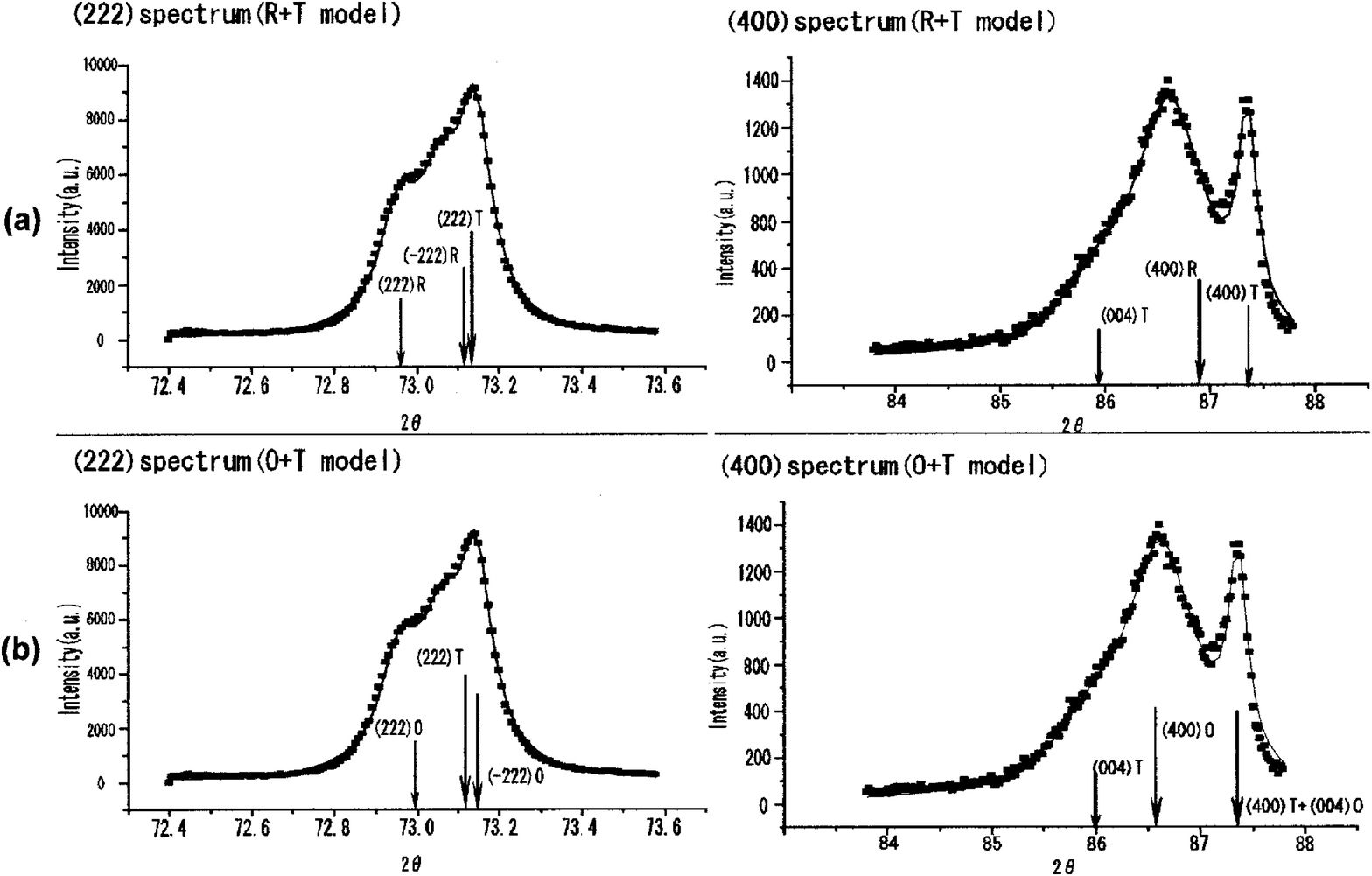}}
\caption{
Diffraction profiles of (222) and (400) reflections of 9PT-1 
measured by X-ray diffraction method (CuK$\beta, \lambda$=1.3922\AA). 
Lines indicate expected diffraction positions of rhombohedral-tetragonal 
mixture model(upper figure), and of orthorhombic-tetragonal mixture 
model(lower figure).
}
\label{fig:6}
\end{fullfigure}

\subsection{Comparison of four 9PT samples}
We also determined lattice parameters of 9PT-4 by neutron diffraction method by simple scanning along [h00] and [0k0] directions. 9PT-4 is a (001) single crystal plate, which has been under the electric field of maximum 10kV/cm. It was found that the symmetry of 9PT-4 is orthorhombic with same lattice parameters as 9PT-2.

The lattice parameters of 9PT-1, 2, 4, and 5 are tabulated in 
Table~\ref{table:1}, and their temperature dependences are shown 
in Fig.~\ref{fig:5}.
These results indicate clearly that the ground state of PZN/9PT, 
the morphotropic phase boundary compound between PZN and ferroelectric 
PT, is orthorhombic Amm2 (9PT-1, 2, 4) with spontaneous 
polarization along the [101] axis or monoclinic Pm with polarization 
vector lying in the (010) plane (9PT-5). 
The space group in orthorhombic state is Amm2 with Ps along the 
[011] direction, similar to the orthorhombic phase in 
BaTiO$_{3}$,~\cite{rf:12} while that of the monoclinic phase 
is Pm judging from the domain configuration shown 
in Fig.~\ref{fig:3}(b) and ~\ref{fig:4}, and the fact that 
monoclinic lattice constants are smoothly connected with those 
of orthorhombic constants with increasing temperature from 10K.
The reason of this 
difference of symmetry could be explained by a slight change 
of composition ratio of PT to PZN, or by the existence of small 
local field due to internal stress, impurity, etc, as the energy 
surface near the symmetric orthorhombic axis is expected to be flat 
(see the part II of the paper).~\cite{rf:15}

In conclusion, the thermal equilibrium state of PZN/9PT at room temperature is not the mixture of rhombohedral and tetragonal phases of end members of pure PZN and PT, but orthorhombic or monoclinic single phase. This symmetry-lowering at the MPB has been already found in PZT48 and PZN-8PT, and the present result confirms that the phenomenon is quite common in Pb-based MPB compounds. 
However, it should be stressed that the present study disclosed a new 
polarization rotation path M$_{\mathrm{c}}^{\#}$ 
(orthorhombic-monoclinic-tetragonal) 
in PZN-9PT, which is different from M$_{\mathrm{a}}$ 
in PZT48 and R-M$_{\mathrm{a}}$-M$_{\mathrm{c}}$ 
in PZN-8PT.~\cite{rf:16}
It was also found that the intrinsic symmetry at the MPB could be obtained 
after an on-off cycle of electric field along the [001] direction, or a 
cooling-heating cycle. 

A theoretical treatment of the present results is described in Part II 
of the paper.~\cite{rf:15}

\section*{Acknowledgements}
We are grateful to Dr. Y. Yamashita, R \& D center of Toshiba Co., 
for providing us PZN/9PT samples, and Dr. J. M. Kiat, SPMS, Ecole 
Centrale Paris, for his cooperation in X-ray diffraction experiment 
of 9PT-1.

Financial supports of Grant-In-Aid for Science Research from 
Monbu-Kagakusho, Grant for Development of New Technology from 
Shigaku-Shinkozaidan , Waseda University Grant for Special 
Research Projects and US DOE under contract No.AC0298CH10866 
are also gratefully acknowledged. This work was performed under 
US-Japan Cooperative Neutron Research Program.


\begin{thebibliography}{99}
\bibitem{rf:1} S-E. Park and T. R. Shrout: 
J. Appl. Phys. {\bf 82} (1997) 1804.
\bibitem{rf:2} J. Kuwata, K. Uchino and S. Nomura: 
Ferroelectrics {\bf 37} (1981) 579.
\bibitem{rf:3} B. Jaffe, W. R. Cook, H. Jaffe: 
{\it Piezoelectric Ceramics} (Academic Press, London, 1971).
\bibitem{rf:4} H. Fu and R. E. Cohen: 
Nature {\bf 403} (2000) 281. 
\bibitem{rf:5} D. Vanderbilt and M. H. Cohen: 
Phys. Rev. B {\bf 63} (2001) 94108.
\bibitem{rf:6} L. Bellaiche, A. Garcia and D. Vanderbilt: 
Phys. Rev. Lett. {\bf 84} (2000) 5427.
\bibitem{rf:7} B.Noheda, D. E. Cox, G. Shirane, E. Cross and S-E. Park: 
Appl. Phys. Lett. {\bf 74} (1999) 2059.
\bibitem{rf:8} B. Noheda, J. A. Gonzalo, E. Cross, R. Guo, S-E. Park, 
D. E. Cox and G. Shirane: 
Phys. Rev. B {\bf 61} (1999) 8687.
\bibitem{rf:9} R. Guo, E. Cross, S-E. Cross, B. Noheda, D. E. Cox and 
G. Shirane: 
Phys. Rev. Lett. {\bf 84} (2000) 5423.
\bibitem{rf:10} B. Noheda, D. E. Cox, G. Shirane, R. Guo, B. Jones and 
E. Cross: Phys. Rev. B {\bf 63} (2001) 14103. 
\bibitem{rf:11} B. Noheda, D. E. Cox, G. Shirane, S-E. Park, E. Cross 
and Z. Zhong: Phys. Rev. Lett. {\bf 86} (2001) 3891.
\bibitem{rf:12} D. E. Cox, B. Noheda, G. Shirane, Y. Uesu, K. Fujishiro 
and Y. Yamada: Appl.Phys.Lett. (9th issue of July, 2001 in press).
\bibitem{rf:13} M. L. Mulvihill, S-E. Park, G. Risch, Z. Li, K. Uchino 
and T. R. Shrout: Jpn. J. Appl. Phys. {\bf 35} (1996) 3984.
\bibitem{rf:14} Y. Uesu, Y. Yamada, K. Fujishiro, H. Tazawa, S. Enokido, 
J. M. Kiat and B. Dkhil: Ferroelectrics {\bf 217} (1998) 319.
\bibitem{rf:15} Y. Yamada, Y. Uesu, M. Matsuda, K. Fujishiro, D. E. Cox, 
B. Noheda and G. Shirane: This issue. 
\bibitem{rf:16} K. Ohwada, K. Hirota, P. W. Rehrig, P. M. Gehring, B. Noheda, 
Y. Fujii, S-E. Park and G. Shirane:  J. Phys. Soc. Japan (condense matter 
\#0105086) (submitted).
\end{thebibliography}
\end{document}